
\let\includefigures=\iftrue
%
\let\includefigures=\iffalse
%
\let\useblackboard=\iftrue
%
%
%
\input harvmac.tex
\message{If you do not have epsf.tex (to include figures),}
\message{change the option at the top of the tex file.}
\input epsf
\epsfverbosetrue
\def\fig#1#2{\topinsert\epsffile{#1}\noindent{#2}\endinsert}
\def\fig#1#2{}
%
\def\Title#1#2{\rightline{#1}
\ifx\answ\bigans\nopagenumbers\pageno0\vskip1in%
\baselineskip 15pt plus 1pt minus 1pt
\else
\def\listrefs{\footatend\vskip 1in\immediate\closeout\rfile\writestoppt
\baselineskip=14pt\centerline{{\bf References}}\bigskip{\frenchspacing%
\parindent=20pt\escapechar=` \input
refs.tmp\vfill\eject}\nonfrenchspacing}
\pageno1\vskip.8in\fi \centerline{\titlefont #2}\vskip .5in}

\ifx\answ\bigans\def\tcbreak#1{}\else\def\tcbreak#1{\cr&{#1}}\fi
\useblackboard
\message{If you do not have msbm (blackboard bold) fonts,}
\message{change the option at the top of the tex file.}
\font\blackboard=msbm10 scaled \magstep1
\font\blackboards=msbm7
\font\blackboardss=msbm5
\newfam\black
\textfont\black=\blackboard
\scriptfont\black=\blackboards
\scriptscriptfont\black=\blackboardss

\else

\fi
%
\def\yboxit#1#2{\vbox{\hrule height #1 \hbox{\vrule width #1
\vbox{#2}\vrule width #1 }\hrule height #1 }}
\def\fillbox#1{\hbox to #1{\vbox to #1{\vfil}\hfil}}
\def\ybox{{\lower 1.3pt \yboxit{0.4pt}{\fillbox{8pt}}\hskip-0.2pt}}
\def\comments#1{}

\def\p{\partial}

\def\tr{{\rm tr\ }}

\def\ket#1{|#1\rangle}

\Title{\vbox{\baselineskip12pt
\hfill{\vbox{
\hbox{EFI-97-29\hfil}
\hbox{UU-HEP/97-3}}}}}
{\vbox{\centerline{Master Ward Identity
for Nonlocal Symmetries}
\vskip20pt
\centerline{in D=2 Principal Chiral Models}}}
\centerline{Miao Li$^1$  and Yong-Shi Wu$^2$}
\smallskip
\centerline{$^1$ Enrico Fermi Institute and Department of Physics}
\centerline{University of Chicago, 5640 S Ellis Ave., Chicago, IL 60637}
\centerline{\tt li@yukawa.uchicago.edu}
\vskip8pt
\smallskip
\centerline{$^2$ Physics Department, University of Utah}
\centerline{Salt Lake City, Utah 84112}
\centerline{\tt wu@mail.physics.utah.edu}
\bigskip
\noindent
\bigskip
\centerline{Abstract}

We derive, in path integral approach, the (anomalous) 
master Ward identity associated with an infinite 
set of nonlocal conservation laws in two-dimensional 
principal chiral models. 

\Date{June 1997}
\nref\lp{M. L\"uscher and K. Pohlmeyer, Nucl. Phys. B137 (1978) 46.}
\nref\cz{E. Br\'ezin, C. Itzykson, J. Zinn-Justin and J.B. Zuber,
Phys. Lett. B82 (1979) 442; T. L. Curtright and C. K. Zachos, Phys. 
Rev. D21 (1980) 411.}
\nref\hgw{B. Y. Hou, M. L. Ge and Y. S. Wu, Phys. Rev. D24 (1981) 2238.}
\nref\dolan{L. Dolan, Phys. Rev. Lett. 47 (1981) 1371.}
\nref\gw{M.L. Ge and Y.S. Wu, Phys. Lett. B108 (1982) 411.}
\nref\df{C. Devchand and D. B. Fairlie, Nucl. Phys. B194 (1982) 232.}
\nref\ys{Y. S. Wu, Nucl. Phys. B211 (1983) 160.}
\nref\bakas{I. Bakas, Nucl. Phys. B428 (1994) 374; Phys. Lett. B343
(1995) 103.}
\nref\maha{J. Maharana, Phys. Rev. Lett. 75 (1995) 205, hep-th/9502001; 
Mod. Phys. Lett. A11 (1995) 9, hep-th/9502002.}
\nref\schwarz{J. H. Schwarz, Nucl. Phys. B447 (1995) 137, hep-th/9504090.}
\nref\sen{A. Sen, Nucl. Phys. B447 (1995) 62, hep-th/9503057.}
\nref\js{J. H. Schwarz, Nucl. Phys. B454 (1995) 427, hep-th/9506076.}
\nref\bs{T. Banks and L. Susskind, Phys. rev. D54 (1996) 1677, 
hep-th/9511193.}
\nref\geroch{R. Geroch, J. Math. Phys. 13 (1972) 394; W.~Kinnersley,
J. Math. Phys. 18 (1977) 1529; V.~Belinskii and V.~Zakharov, Zh. Eksp.
Teor. Fiz. 75 (1978) 1955; H.~Nicolai, Phys. Lett. B194 (1987) 402;
 P.~Breitenlohner and 
D.~Maison, Ann. Poincare 46 (1987) 215.}
\nref\luscher{M. L\"uscher, Nucl. Phys. B135 (1978) 1.}
\nref\abda{E. Abdalla, M. Gomes and M. Forger, Nucl. Phys. B210
(1982) 181; E.~Abdalla, M.~C.~B.~Abdalla and M.~Forger, Nucl. Phys.
B297 (1988) 374.} 
\nref\antal{B. E. Fridling and A. Jevicki, Phys. Lett. B134 (1984) 70.}
\nref\fujikawa{K. Fujikawa, Phys. Rev. Lett 42 (1979) 1195; Phys.
Rev. D21 (1980) 2848; Phys. Rev. Lett. 44 (1980) 1733.} 
\nref\poly{A. M. Polyakov, ``Gauge Fields and Strings'', 
Harwood Academic Publishers (1987).}
\nref\bernard{D. Bernard, Comm. Math. Phys. 137 (1991) 191.}
\nref\fkw{V. A. Fateev, V. A. Kazakov and P. B. Wiegmann, Nucl. Phys.
B424 (1994) 505.}

In a meeting with one of the present authors (Y.S.W.), 
Prof. Namiki made the statement that (one of) the main 
theme(s) of his own researches is fluctuations, including 
both the statistical and quantum ones and their interplay. 
It is well-known that one effect of quantum fluctuations
is their modification (or even destruction) of classical 
symmetries, under the name of anomalies. Here we would 
like to devote to this volume commemorating Prof. Namiki
the present note on the master Ward identity that governs
the fate in the quantum theory of the duality symmetries
associated with an infinite set of nonlocal conservation laws 
in $D=2$ principal chiral models.

Originally the $d=2$ nonlinear sigma models, including 
principal chiral models (PCM), attracted a lot of attention 
because of the recognition of many similarities between 
them and $D=4$ non-Abelian gauge theories. An infinite 
set of nonlocal conservation laws were uncovered in PCM
about twenty  years ago \lp. The associated nonlocal 
charges generate symmetries of the equations of motion, 
producing new solutions from an old solution. The 
off-shell version of such symmetries was discovered in \hgw, 
where the nonlocal transformations are summarized by 
a one-parameter family of  transformations, and  the variation 
of the Lagrangian under them  is shown to be a 
total divergence. The infinite-dimensional symmetry algebra 
was derived by a number of authors \refs{\dolan,\gw,\df,\ys}.

Recently, interest in such symmetries is revived,  as they may 
give rise to duality symmetries in dimensionally reduced string 
theory \refs{\bakas-\js}. When String/M theory is compactified 
to a lower dimension, the duality symmetry becomes very rich. 
There are signs indicating that more degrees of freedom appear 
in lower dimensions. However, 2D string theory is peculiar. In 
the low energy sector, a 2D model coupled to dilaton gravity is 
obtained. This 2D model is a certain coset model and is closely 
related to PCM. The coset space is not really the moduli space 
of vacuum, since the measure on this space is compact, and the 
special 2D infrared kinematics dictates that the whole moduli 
space is explored by a quantum state. In higher dimensions, duality 
symmetry often maps one description of the theory to another. 
In two dimensions, we expect duality symmetry plays a more 
dynamical role, and it is not clear whether the real ''moduli space"
is a further quotient by a discrete duality group. A discussion 
on quantum degrees of freedom in critical 2D string theories has
been given in \bs. A sample of references on nonlocal symmetries 
in PCM coupled to gravity is given in \geroch, where our listing 
is far from complete. 

The fate of the nonlocal symmetries in the quantum theory
is of great interest. Previously the first quantum non-local 
charge was defined by L\"uscher for the $O(N)$ non-linear sigma 
model \luscher\ and used to deduce the factorized S-matrix in
the canonical approach, which uses the equations of motion in
operator form. (Generalization to PCM and related models 
was done in \abda.) Needless to say, a systematic understanding 
of quantum aspects of all nonlocal charges  will deepen our 
understanding of PCM and may even help to solve these models. 
Also, the issue of whether the large symmetry is discretized at 
quantum level (as in higher dimensional string theories) will be 
settled only after a systematic knowledge of quantum corrections 
become available. Despite significant progress made in 
understanding duality in higher dimensional string theories, 
little is known about quantum duality symmetry in three and two 
dimensions. The present work is an effort toward that direction;
we will work in path integral approach and derive the master
Ward identity,  which contains the spectral parameter and 
summarizes all nonlocal charges. 

In a quantum theory, it is the Ward identities for Green's functions 
that fully describe the dynamical effects of quantum fluctuations on 
the symmetry of the action. In addition to quantum mechanical 
transformation of composite operators, the Ward identities indicate 
whether the symmetry is anomalous. If it is, the local Ward identities 
will involve an additional term corresponding to insertion of the 
anomalous correction to the current. The well-know examples 
are the chiral anomaly in gauge theory, and conformal anomaly 
in conformal field theories. If such an anomaly exists, then most 
likely the symmetry algebra is corrected quantum mechanically;
again a well-known example is the conformal anomaly and 
the associated central charge in the Virasoro algebra.
We will find that indeed the nonlocal symmetries of  PCM's
are anomalous, and the finite anomaly part is to be computed 
in this paper. For reason we will present shortly, the Noether
currents corresponding to these symmetries are not realized in
the conventional way, therefore the anomaly in the symmetry 
algebra can not be deduced straightforwardly. We leave this 
problem to future study.

We work in two dimensional Euclidean space in this paper, the
corresponding result in Minkowski space is readily obtained 
by proper Wick rotation. The group is chosen to be $SU(N)$, 
though it is straightforward to generalize our result to other 
groups and coset spaces. The classical action of the PCM is
\eqn\action{S={1\over 2g^2}\int d^2x\, \tr A^2_\mu,}
where $A_\mu$ are traceless Hermitian matrices of the form
$A_\mu=-iG^{-1}\p_\mu G$, $G$ taking values 
in $SU(N)$. Alternatively, one imposes the flatness condition 
on $F_{01}=\p_0A_1-\p_1A_0+i[A_0,A_1] =0$. The 
equation of motion reads $\p_\mu A_\mu=0$. Classically, these
two equations fully determine the theory. It turns out that 
these two equations are equivalent to the following Lax pair \cz:
\eqn\lax{\eqalign{\p_1U&={il\over 1+l^2}\left(A_0-lA_1\right)U,\cr 
\p_0U&=-{il\over 1+l^2}\left(A_1+lA_0\right)U,}}
where $U$ is a unitary matrix, and $l$ is a real parameter. 
The Minkowski version is obtained by Wick
rotation: $x_0\rightarrow ix_0$, $A_0\rightarrow -iA_0$,
$l \rightarrow -il$. It is easy to check that the equation of motion
is invariant under the transformation 
\eqn\transf{\delta_\epsilon A_\mu=D_\mu\left(U(x)\epsilon U^{-1}(x)\right),}
where $\epsilon$ is a constant Hermitian matrix, and the covariant 
derivative is given by $D_\mu=\p_\mu+i[A_\mu,\cdot]$. It was shown
by a number of authors \refs{\dolan,\gw,\df} that upon Taylor expanding
$U$ in $l$, a half of Kac-Moody algebra arises from commutators 
of these symmetries, which is later enlarged to the full Kac-Moody 
algebra by one of us \ys, by including also generators obtained by 
expanding $U$ in Taylor series in $l^{-1}$, whose importance was
emphasized recently by Schwarz \schwarz  in the context of duality 
symmetries.

In this paper we are going to use the path integral approach, in 
which quantum fluctuations are represented by field configurations
that do not satisfy classical equations of motion. Therefore one of 
the equations in \lax\ must be abandoned. Without loss of generality, 
we keep the first one as the defining equation for $U$ and 
solve $U$ as follows
\eqn\defi{U(x)=\overleftarrow{P}\exp\left({il\over 1+l^2}\int^{x_1}
_{-\infty}(A_0-lA_1)(y)dy_1\right).}
For later use, we also define $U(x,y)=U(x)U^{-1}(y)$. It was 
shown in \hgw\ that the action is invariant up to a total divergence
under the transformation \transf\
with the above off-shell definition of $U$. To derive the local
Ward identities, we need to know the variation of the action with
a function $\epsilon(x)$, not just a constant. The variation is
simply
\eqn\fvari{\delta S=-{1\over g^2}\int d^2x\, 
\tr U^{-1}(\p_\mu A_\mu) U\epsilon(x),}
(A total divergence is discarded, since the surface term is always
zero by properly choosing $\epsilon(x)$.)
and upon using the flatness condition and the definition in \defi,
the variation is written in a form \hgw
\eqn\svari{\delta S={1\over g^2}\int d^2x\,
\tr(\p_\mu J_\mu)\epsilon(x),}
with the current
\eqn\curre{J_\mu=\epsilon_{\mu\nu}\left[lU^{-1}A_\nu U-i(l+l^{-1})
U^{-1}\p_\nu U\right].}
We see that $\delta S$ is vanishing off-shell up to a total divergence,
when $\epsilon$ is a constant. One is tempted to
conclude that the above current is the Noether current, since it is 
conserved on-shell. Unfortunately, it is easy to see that $J_0=0$ 
identically, due to the definition \defi. Moreover $J_1=0$ on-shell. 
Several non-vanishing conserved currents are found in \hgw. 
Nevertheless, it is $J_1$ which will appear in local Ward identities. 
One certainly can not set $J_1=0$ in Green's functions. Equating 
\fvari\ and \svari, we have
\eqn\eqal{\p_1J_1=-U^{-1}(\p_\mu A_\mu)U,}
so the divergence of $J_1$ is just an adjoint transformation of 
the equations of motion.

To begin our calculation, a convenient definition of the path
integral is needed. It will be shown that the following path integral
\eqn\pathi{\langle F\rangle=N^{-1}\int [dA_\mu]\delta(F_{01})Fe^{-S}}
is equivalent to the conventional one, where the action is given in
\action. The delta function factor reduces the path integral to the
sub-space of flat connections. This delta function can be replaced
by introducing a Lagrange multiplier field $B$. The $B$ field is sort 
of dual formulation of the principal chiral model, as was studied in
\antal, where it was shown that the beta function for $g^2$ calculated
in this dual formulation is the same as calculated in the original 
formulation. Keeping both $B$ and $A$ will prove convenient 
for a polynomial canonical formulation which we plan to study 
in the future. At present, we wish to demonstrate that \pathi\ 
is equivalent to a conventional path integral. It is enough to show 
this is true locally, so let us consider the neighborhood of 
a flat connection $\hat{A}_\mu$. Write $A_\mu=\hat{A}_\mu 
+a_\mu$, $a_\mu$ is the fluctuation. The measure is defined 
according to the norm $|a|^2=\int d^2x\, \tr a_\mu^2$. The field 
strength, to the first order, reduces to $F_{01}=D_0a_1-D_1a_0$, where 
the covariant derivative is defined with $\hat{A}_\mu$. Decompose 
$a_\mu=D_\mu\phi+\epsilon_{\mu\nu}D_\nu\psi$, $\phi$ and $\psi$ 
both are a Hermitian matrix field. Due to the flatness of 
$\hat{A}_\mu$, the norm undergoes an orthogonal decomposition 
$|a|^2=\int d^2x\, \tr \left((D_\mu\phi)^2 +(D_\mu\psi)^2\right)$. 
Furthermore, the fluctuation represented by $\psi$ is orthogonal 
to the subspace of flat connections, since $F_{01}=-{1\over 2} D^2\psi$. 
To finish our argument, note that by changing integration variables 
from $A_\mu$ to $\phi$ and $\psi$, the resulting Jacobian is 
$\det(-D^2)$, which gets cancelled by a factor 
from the delta function in integrating out $\psi$. 

We adopt Fujikawa's path integral method to compute the anomaly
\fujikawa. Under the transformation $\tilde{A}=A+\delta A$ with
$\delta A$ given by \transf\ with a function $\epsilon(x)$ in that
formula, the whole quantity \pathi\ remains unchanged, since changing
the integration variable does not change the result. Thus, an 
anomalous Ward identity results:
\eqn\ward{\delta \left(\langle F\rangle\right)= \langle \delta F\rangle-
\langle \delta S\rangle +\langle\delta\det\rangle=0,}
where $\delta F$ is the change in $F$ under the transformation. If
$F$ is a composite operator, this transformation is subject to 
renormalization effects. $\delta S$ is given in \svari, and $\delta\det$ 
is the change in the measure
$$\delta\det=\delta\det\left({\p (\tilde{A})\over \p (A)}\right).$$
Notice that there is no change brought about by the delta function factor,
since $F_{01}$ transforms as the adjoint representation, and the 
delta function remains invariant. The above expression is usually 
divergent, and a proper regularization is needed. 

According to our discussion before, we expand
\eqn\expan{a_\mu=\sum_n\left(a_nD_\mu\phi_n+b_n\epsilon_{\mu\nu}
D_\nu\phi_n\right),}
with the eigen-vector equation
\eqn\eigen{-D^2\phi_n=\lambda_n\phi_n,}
and the orthonormality condition 
$\int d^2x\, \tr D_\mu\phi^+_nD_\mu\phi_m=\delta_{nm}$, 
or, using the eigen-value equation $\int d^2x\, \tr
\phi^+_n\phi_m=\lambda_n^{-1}\delta_{nm}$.
Now the measure $[dA_\mu]$ is defined by 
$\prod_n[da_ndb_n]$, and the change of measure is given by
\eqn\chan{\delta\det =\sum_n({\p\delta a_n\over \p a_n}+
{\p\delta b_n\over \p b_n})e^{-\lambda_n t},}
this being already regularized \fujikawa. Now
$$\eqalign{{\p\delta a_n\over \p a_n}&=\int d^2xd^2y{\p\delta A^{ij}_\mu
(x)\over \p A^{lk}_\nu(y)}D_\mu\phi^{+ji}_n(x)D_\nu\phi^{lk}_n(y),\cr
{\p\delta b_n\over \p b_n}&=\int d^2xd^2y{\p\delta A^{ij}_\mu
(x)\over \p A^{lk}_\nu(y)}\epsilon_{\mu\lambda}\epsilon_{\nu\sigma}
D_\lambda\phi^{+ji}_n(x)D_\sigma\phi^{lk}_n(y),}$$
where the sum over all repeated indices except for $n$ is assumed. 
Substituting the above expressions into \chan\ we obtain
\eqn\schan{\delta\det =\int d^2xd^2y{\p\delta A^{ij}_\mu
(x)\over \p A^{lk}_\nu(y)}K^{\mu\nu}_{ji,lk}(x,y,t),}
with the heat-kernel
\eqn\heat{K^{\mu\nu}(x,y,t)=\left(D_\mu(x)D_\nu(y)+\epsilon_{\mu
\lambda}\epsilon_{\nu\sigma}D_\lambda(x)D_\sigma(y)\right)
\sum_n\phi^+_n(x)\otimes \phi_n(y)e^{-\lambda_n t},}
where we used the symbol $\otimes$ to remind ourselves that $\phi_n^+(x)$
and $\phi_n(y)$ carry independent matrix indices.

The next step is to compute the heat kernel. To solve the eigen-value
problem \eigen, observe that for a flat connection $A_\mu=-iG^{-1}\p_\mu
G$, $D_\mu\phi_n=G^{-1}\p_\mu(G\phi_n G^{-1})G$ and $-D^2\phi_n
=-G^{-1}\p^2(G\phi_nG^{-1})G$. So the eigen-value equation reads 
$$-\p^2(G\phi_nG^{-1})=\lambda_nG\phi_nG^{-1},$$
reducing to the eigen-value problem without connection. Let the scalar
eigen-function $\phi_i$ be the one satisfying $-\p^2\phi_i
=\lambda_i\phi_i$, then an eigen-function $\phi_n$ can be written as 
\eqn\solv{\phi_n(x)=\phi_i(x)G^{-1}(x)T^aG(x),\quad \lambda_n=\lambda_i,}
where $T^a$ is a generator of the $su(N)$ algebra, a traceless Hermitian
matrix. To satisfy the normalization condition, one then imposes
$\tr T^aT^b=\delta_{ab}$ and $\int d^2x\bar{\phi}_i\phi_j=\lambda_i^{-1}
\delta_{ij}$. With the result \solv, we have
\eqn\tens{\sum_n\phi^+_n(x)\otimes \phi_n(y)e^{-\lambda_n t}
=\sum_a [G^{-1}(x)T^aG(x)]\otimes [G^{-1}(y)T^aG(y)]\sum_i
\bar{\phi}_i(x)\phi_i(y)e^{-\lambda_i t}.}
The last factor can be written in a continuum form
$$\sum_i
\bar{\phi}_i(x)\phi_i(y)e^{-\lambda_i t}=\int {d^2k\over (2\pi)^2}
k^{-2}e^{-k^2t+ik(x-y)}.$$
To obtain the heat kernel, substitute \tens\ into \heat\ and notice
the fact that 
$$D_\mu(x)[G^{-1}(x)T^aG(x)]=D_\mu(y)[G^{-1}(y)T^aG(y)]=0,$$
we find
\eqn\heatk{K^{\mu\nu}(x,y,t)={\delta_{\mu\nu}\over 4\pi t}
\exp\left(-{(x-y)^2\over 4t}\right)\sum_a [G^{-1}(x)T^aG(x)]\otimes 
[G^{-1}(y)T^aG(y)],}
where we used
$$\eqalign{&(-\p_\mu\p_\nu-\epsilon_{\mu\lambda}\epsilon_{\nu\sigma}
\p_\lambda \p_\sigma)\int {d^2k\over (2\pi)^2}k^{-2}e^{-k^2t+ik(x-y)}\cr
&=
\int {d^2k\over (2\pi)^2}k^{-2}(k_\mu k_\nu+\epsilon_{\mu\lambda}
\epsilon_{\nu\sigma}k_\lambda k_\sigma)e^{-k^2t+ik(x-y)}\cr
&=\delta_{\mu\nu}\int {d^2k\over (2\pi)^2}e^{-k^2t+ik(x-y)}
={\delta_{\mu\nu}\over 4\pi t}\exp\left(-{(x-y)^2\over 4t}\right).}$$
Substituting the heat kernel \heatk\ into \schan, the regularized
change in the measure is
\eqn\tchan{\delta\det={1\over 4\pi t}\int d^2xd^2y\exp\left(-{(x-y)^2
\over 4t}\right){\p\delta A^{ij}_\mu (x)\over \p A^{lk}_\mu(y)}
[G^{-1}(x)T^aG(x)]^{ji}[G^{-1}(y)T^aG(y)]^{lk},}
again with the sum over repeated indices assumed.

To complete our calculation, we now endeavor to calculate 
\eqn\fder{{\p\delta A^{ij}_\mu (x)\over \p A^{lk}_\mu(y)}.}
To this end, we need the following variation formula
\eqn\variat{d\left(U(x)\epsilon(x)U^{-1}(x)\right)
={il\over 1+l^2}\int^{x_1}_{-\infty}dy_1[U(x,y)(dA_0-ldA_1)(y)
U(y)\epsilon(x)U^{-1}(x)-h.c.],}
where $U(x,y)=U(x)U^{-1}(y)$. We let $\epsilon(x)$ be
a function, in order to derive local anomalous Ward identities.
To compute functional derivatives, first notice that from \transf
$$d\delta A_\mu(x)=D_\mu\left(d(U(x)\epsilon(x)U^{-1}(x))\right)
+i[d A_\mu(x),U(x)\epsilon(x) U^{-1}(x)].$$
Using this formula and \variat\ we derive, after a little lengthy
calculation,
\eqn\zero{\eqalign{&{\p\delta A^{ij}_0 (x)\over \p A^{lk}_0(y)}
[G^{-1}(x)T^aG(x)]^{ji}[G^{-1}(y)T^aG(y)]^{lk}\cr
&={il\over 1+l^2}\p_0[\theta(x_1-y_1)\delta(x_0-y_0)\tr\left(
\tilde{U}(y)\epsilon(x)\tilde{U}^{-1}(x)T^a\tilde{U}(x,y)T^a-h.c.\right)]\cr
&+i\delta^2(x-y)\tr\left( G^{-1}(x)T^aG(x)[G^{-1}(y)T^aG(y),U(x)\epsilon(x)
U^{-1}]\right),}}
where $\theta$ is the step function, and $\tilde{U}(x)=G(x)U(x)$,
$\tilde{U}(x,y)=\tilde{U}(x)\tilde{U}^{-1}(y)$. The additional factor $G$
comes from the factor in the heat kernel. Next, we demonstrate that
this term does not contribute to $\delta\det$ after substitution into
\tchan. First, consider the contribution of the first term on the R.H.S.
of \zero:
$$\eqalign{&{1\over 4\pi t}\int d^2xd^2y\p_0[\theta(x_1-y_1)\delta(x_0-y_0)
(\cdots)]\exp\left(-{(x-y)^2\over 4t}\right) \cr
&={1\over 4\pi t}\int d^2xd^2y\theta(x_1-y_1)\delta(x_0-y_0) (\cdots)
{x_0-y_0\over
2t}\exp\left(-{(x-y)^2\over 4t}\right)=0,}$$
where we integrated by parts with respect to $x_0$. Next consider 
the contribution of the second term on the R.H.S. of \zero:
$$\eqalign{&{i\over 4\pi t}\int d^2xd^2y \delta^2(x-y)(\cdots)
\exp\left(-{(x-y)^2\over 4t}\right) \cr
=&{i\over 4\pi t}\int d^2x\tr\left(G^{-1}(x)T^aG(x)[G^{-1}(x)T^aG(x),
U(x)\epsilon(x)U^{-1}(x)]\right)=0.}$$

Similar to \zero, one derives
\eqn\first{\eqalign{&{\p\delta A^{ij}_1 (x)\over \p A^{lk}_1(y)}
[G^{-1}(x)T^aG(x)]^{ji}[G^{-1}(y)T^aG(y)]^{lk}\cr
&=-{il^2\over 1+l^2}\p_0[\theta(x_1-y_1)\delta(x_0-y_0)\tr\left(
\tilde{U}(y)\epsilon(x)\tilde{U}^{-1}(x)T^a\tilde{U}(x,y)T^a-h.c.\right)]\cr
&+i\delta^2(x-y)\tr\left( G^{-1}(x)T^aG(x)[G^{-1}(y)T^aG(y),U(x)\epsilon(x)
U^{-1}]\right).}}
Again, the second term on the R.S.H. of \first\ does not contribute
to $\delta\det$. Thus, the only non-vanishing contribution comes from the
first term on the R.H.S. of \first. This term can be further simplified
by using the following formula
$$\sum_a\tr\left(AT^aBT^a\right)=(\tr A)(\tr B) -{1\over N}\tr (AB),$$
valid for $SU(N)$. Substituting the first term in \first\ into \tchan,
\eqn\fchan{\eqalign{\delta\det&=-{il^2\over (1+l^2)4\pi t}\int d^2xd^2y
\theta(x_1-y_1)\delta(x_0-y_0){x_1-y_1\over 2t}
\exp\left(-{(x-y)^2\over 4t}\right)f(x,y)\cr
&=-{il^2\over (1+l^2)4\pi t}\int d^2x\int^{x_1}_{-\infty}dy_1 
{x_1-y_1\over 2t}\exp\left(-{(x_1-y_1)^2\over 4t}\right)f(x,y),}}
with
\eqn\ket{f(x,y)=\tr[\tilde{U}(x)\epsilon(x)\tilde{U}(y)]\tr
\tilde{U}(x,y)-c.c. .}
Let $x_1-y_1=z\sqrt{t}$ in \fchan, and expand $f(x,x_1-z\sqrt{t})$
to the order $t$, we obtain the singular terms as well as a finite
term 
\eqn\sichan{\eqalign{\delta\det &=-{il^2\over 8\pi(1+l^2)}\left[{2\over t}
\int d^2xf(x,x)-2\sqrt{{\pi\over t}}\int d^2x\p_{y_1}f(x,x)
+4\int d^2x\p^2_{y_1}f(x,x)\right]\cr
&+O(\sqrt{t}).}}
The term proportional to $1/t$ in \sichan\ is absent, since $\tr\tilde{U}
(x)\epsilon \tilde{U}^{-1}(x)=0$ and hence $f(x,x)=0$. The second term is 
nonzero because
$$\p_{y_1}f(x,x)=N\tr [\tilde{U}^{-1}(x)\p_1\tilde{U}(x)  \epsilon(x)]
-c.c.=2N\tr[ \tilde{U}^{-1}\p_1\tilde{U} \epsilon ].$$
So there is a divergent term proportional to $1/\sqrt{t}$. Indeed
we discovered anomaly by first computing this ``bare'' term. Finally,
the third term, being finite, is given by 
$$\eqalign{\p^2_{y_1}f(x,x)&=N\tr[\tilde{U}^{-1}(x)\p_1^2\tilde{U}(x)  
\epsilon(x)]-c.c.\cr
&=N\tr[\p_1\left(\tilde{U}^{-1}(x)\p_1\tilde{U}(x)\right)\epsilon(x)]-c.c.\cr
&=2N\tr[\p_1\left(\tilde{U}^{-1}(x)\p_1\tilde{U}(x)\right))\epsilon(x)]\cr
&={2iN\over 1+l^2}\tr[\p_1\left(U^{-1}(A_1+lA_0)U\right)\epsilon(x)],
}$$
where in the second line we used the fact that $\tr[\p_1\tilde{U}^{-1}
\p_1\tilde{U}\epsilon]$ is real, and in the fourth line we used the
definition of $\tilde{U}=GU$ and the defining equation for $U$, the
first equation in \lax. Plugging the last line into \sichan\ and dropping
the divergent term, the finite anomaly is then
\eqn\anom{(\delta\det)_R={Nl^2\over \pi (1+l^2)^2}\int d^2x\,
\tr[\p_1\left(U^{-1}(A_1+lA_0)U\right)\epsilon(x)].}
It is not surprising to see that when $\epsilon$ is a constant, 
the anomaly is a total divergence.

With the result \anom\ at hand, we easily write down local anomalous
Ward identities. Come back to \ward\ in which take 
$\epsilon(x)=T^a\alpha(x)$. Taking the functional derivative of 
\ward\ with respect to $\alpha(x)$, we then have
\eqn\anomw{\langle{\delta F\over\delta\alpha(x)}\rangle=
{1\over g^2}\left[\p_1\langle J_1^a(x)F\rangle-\p_1
\langle j^a(x)F\rangle\right],}
where the first term on the R.H.S. comes from $\delta S$, and the 
second term is the anomalous term. Explicitly,
\eqn\current{\eqalign{J_1^a(x)&=\tr[\left(i(l+l^{-1})\p_0U-
(A_1+lA_0)U\right)T^aU^{-1}],\cr
j^a(x)&={g^2Nl^2\over \pi (1+l^2)^2}\tr[(A_1+lA_0)UT^aU^{-1}].}}
It is interesting to observe that both the original current $J_1^a$
and the anomalous current only have the spatial component, and the 
anomalous current modifies the coefficient of the second term of 
the original current in \current. It is obvious that the anomalous 
part is a "one-loop" quantum correction, since it is multiplied by 
$g^2$ compared to $J_1$.

Eq. \anomw\ is the main result of the present note. 
It can be viewed as the "master" Ward identity, since it 
encodes infinitely many Ward identities by expanding $U$ 
in $l$ or $l^{-1}$. To properly understand \anomw, one 
would have to take care of transformation rule for a 
composite operator $F$, in which renormalization 
effects are included. It is well-known that if a conserved 
current is not anomalous, then it is not renormalized. What 
we have learned from our computation is that all those 
infinitely many nonlocal currents are anomalous, except for
the first two upon expanding $U$ in the Taylor series in $l$ 
(or in $l^{-1}$). Let us remind ourselves that $\p_1J_1^a$ 
can also be written as $-\tr[\p_\mu A_\mu UT^aU^{-1}]$, 
according to \eqal. We immediately see that the $l^0$ term 
is just $-\p_\mu A^a_\mu$, this is the first conserved current 
of the infinite set \lp. There is no zeroth order in the quantum 
correction, so this current is anomaly free.
It follows from \anomw\ by taking $F=1$ that
\eqn\resl{\langle \p_\mu A_\mu \rangle=0.}
This agrees with the finding in \abda\ that the local charge is not
renormalized. The term of order $l$ in $\p_1J_1$ is also anomaly 
free. This does not contradicts the nontrivial renormalization of 
the first non-local charge \refs{\luscher,\abda}, since $J_0=0$ in 
our discussion. The fact that $\langle A_\mu \rangle=\langle
-iG^{-1}\p_\mu G \rangle $ is not renormalized 
does not imply that $A_\mu^2$ is also not renormalized. 
In fact, the Lagrangian in \action\ is proportional to this 
operator, and $g^2$ or equivalently $\tr A_\mu^2$ is renormalized
\poly. This fact particularly indicates that much further work is
to be done in order to understand the Ward identities.

In conclusion, we have shown that most of the infinite set of 
nonlocal symmetries in PCM are anomalous, and we have computed the 
finite quantum correction in the master Ward identity. Generalization 
to other groups and symmetric spaces is not difficult. Much further 
work remains to be done. For example, we are yet to understand the 
implications for the conserved currents constructed for instance in 
\hgw. Also the quantum modification of the classical centerless 
Kac-Moody algebra \refs{\dolan, \gw, \df, \ys, \schwarz} due to 
the anomaly we have computed here is yet to be derived. The 
relationship of our quantum corrections to the Yangian algebra
(as the quantum mechanically corrected symmetry algebra) in 
massive integrable models \bernard\ is to be unraveled too. It is 
hoped that a complete understanding of the quantum symmetries 
could lead to a new method of solving PCM for a compact group, 
and shed light on the large $N$ problem as recently studied in \fkw. 
Perhaps the most intriguing is to understand duality symmetry in 
a non-compact coset model based on PCM, in future developments 
along the line we initiated here and the line presented in \refs{\luscher, 
\abda}.

\noindent {\bf Acknowledgments}

The work of Y.S.W. is supported in part by
U.S. NSF grant PHY-9601277. The work of M.L. is 
supported by DOE grant DE-FG02-90ER-40560 
and NSF grant PHY-9123780.

\listrefs

\end